\documentclass[preprint,aps]{revtex4}

\usepackage{graphics}
\usepackage{amsmath}

\begin{document}

\title{The Onset of Chaotic Motion of a Spinning Particle around the Schwarzchild Black Hole}

\author{J.-K. Kao}
  \email{g3180011@tkgis.tku.edu.tw}
\author{H. T. Cho}
  \email{htcho@mail.tku.edu.tw}
\affiliation{Department of Physics, Tamkang University, Tamsui,
Taipei, Taiwan, Republic of China}

\date{\today}

\begin{abstract}

In the Schwarzchild black hole spacetime, we show that chaotic
motion can be triggered by the spin of a particle. Taking the spin
of the particle as a perturbation and using the Melnikov method,
we find that the perturbed stable and unstable orbits are
entangled with each other and that illustrates the onset of
chaotic behavior in the motion of the particle.

\end{abstract}


\maketitle

In astrophysics, the angular momentum or spin of a particle is one
of the most important elements for consideration because nearly
all astrophysical objects possess it. It may also be the crucial
ingredient in the study of chaotic behavior as shown, for example,
by Suzuki and Maeda \cite{suzuki} that the motions of spinning
particles in the Schwarzschild black hole spacetime could be
chaotic by analyzing the Lyapunov exponents \cite{marion}.
Moreover, this is an important result due to its relevancy to the
detection of gravitation waves from black hole coalescences
\cite{levin,cornish} as it may jeopardize the use of matching
templates in interpretating the upcoming data from the LIGO
experiment.

In this Letter we would like to use a different method, called the
Melnikov technique \cite{melnikov}, to investigate the chaotic
motions of spinning particles in the Schwarzschild black hole
spacetime. In \cite{cho}, we have successfully applied this
technique to detect the onset of chaotic motions in the simple
dynamical system of the one-dimensional Duffing hamiltonian with a
spinning particle. In this method one consider homoclinic orbits
emanating and terminating at the same unstable fixed point in an
integrable system. When time-dependent perturbations are added to
the system, the stable and the unstable orbits in the perturbed
system would split in general. The Melnikov function, which is an
integral evaluated along the unperturbed homoclinic orbit,
calculates the transversal distance between the perturbed stable
and unstable orbits on the Poincar$\acute{e}$ section. The isolated zeros in
the Melnikov function indicated complicated entanglements between
the two perturbed orbits and therefore the presence of chaotic
behaviors.

Here we first illustrate that there is a homoclinic orbit in the
two-dimensional phase space of the radial motion of a spinless
particle in the Schwarzschild spacetime. Suppose that the mass of
the Schwarzchild black hole is $M$, and the line element is
described in the usual Schwarzschild coordinates by
\cite{weinberg}
\begin{eqnarray}
ds^2&=&-fdt^2+f^{-1}dr^2+r^2(d\theta^2+\sin^2 \theta
d\phi^2),\nonumber \\ f(r)&=&1-\frac{2M}{r}.
\end{eqnarray}
For a relativistic particle of mass $m$ moving in a curved
background described by the spacetime metric $g_{\mu\nu}$, the
action can be taken to be
\begin{equation}
S[x]=\frac{m}{2}\int g_{\mu\nu} \dot{x}^{\mu} \dot{x}^{\nu}d\tau.
\end{equation}
$x^{\mu}(\tau)$ denotes the spacetime coordinates of the particle
where $\tau$ is the proper time, and we must in addition impose
the mass shell constraint
\begin{eqnarray}
g^{\mu\nu}p_{\mu}p_{\nu}&=&-m^2,\nonumber \\ p_{\mu}&=&mg_{\mu\nu}
\dot{x}^{\nu},\label{masscondition}
\end{eqnarray}
where $p_{\mu}$  is the momentum canonically conjugate to
$x^{\mu}$. Because
 $g_{\mu\nu}$ is stationary and axisysmmetric, the particle has two additional
conserved quantities, the momenta conjugate to $\tau$ and $\phi$,
\begin{eqnarray}
E&=&-p_t=mf\frac{dt}{d\tau},\nonumber \\ L&=&p_{\phi}=mr^2\sin^2
\theta \frac{d\phi}{d\tau}.
\end{eqnarray}
Let us restrict the particle motion, with a fixed angular momentum
$L$, to lie in the $\theta=\pi/2$ plane. Then the only effective
degree of freedom is $r$, and we will find that there is exactly
one homoclinic orbit in the reduced phase space $(r,p_r)$ with
appropriately chosen $E$ and $L$.

Substitute the expressions for $E$ and $L$ into
Eq.~(\ref{masscondition}), a first-order equation of motion for
$r(\tau)$ can be obtained
\begin{eqnarray}
m^2\left(\frac{dr}{d\tau}\right)^2&+&f(r)\left(m^2+\frac{L^2}{r^2}\right)=E^2
\nonumber \\ \Rightarrow
m^2\left(\frac{dr}{d\tau}\right)^2&-&\frac{2Mm^2}{r}+
\frac{L^2}{r^2}-\frac{2ML^2}{r^3}=E^2-m^2.\label{eqofmotion}
\end{eqnarray}
This equation can be interpretated as describing the
one-dimensional problem of a non-relativistic particle with energy
$E^2-m^2$ moving in the potential
\begin{equation}
U_{eff}=-\frac{2Mm^2}{r}+\frac{L^2}{r^2}-\frac{2ML^2}{r^3},\label{ueff}
\end{equation}
which has an attractive $1/r$ term and a repulsive $1/r^2$ term as
in the corresponding Newtonian system, plus an additional
attractive
 $1/r^3$ term which is a purely general relativistic effect. It is the
latter which is responsible for the existence of the homoclinic
orbit.

It is convenient to change the dynamical variable to $u\equiv
2M/r$. The equation of motion in Eq.~(\ref{eqofmotion}) then
becomes \cite{bombelli}
\begin{equation}
(\frac{du}{d\phi})^2-\frac{4M^2m^2}{L^2}u+u^2-u^3=\frac{4M^2(E^2-m^2)}{L^2}.
\end{equation}
It is easy to see that $U_{eff}$ has extrema at real values of
$u$, provided that $12M^2m^2/L^2<1$, and they are located at
\begin{eqnarray}
u_{st}&=&\frac{1}{3}(1-\beta), \nonumber \\
u_{un}&=&\frac{1}{3}(1+\beta).
\end{eqnarray}
where $\beta=\sqrt{1-12M^2m^2/L^2}$ is the parameter in the
problem to characterize the existence and the properties of the
homoclinic orbit. $u_{st}$ is the stable fixed point, while
$u_{un}$ is the unstable one. Figure~1 shows the potential with
$\beta=0.4$.
\begin{figure}
\includegraphics{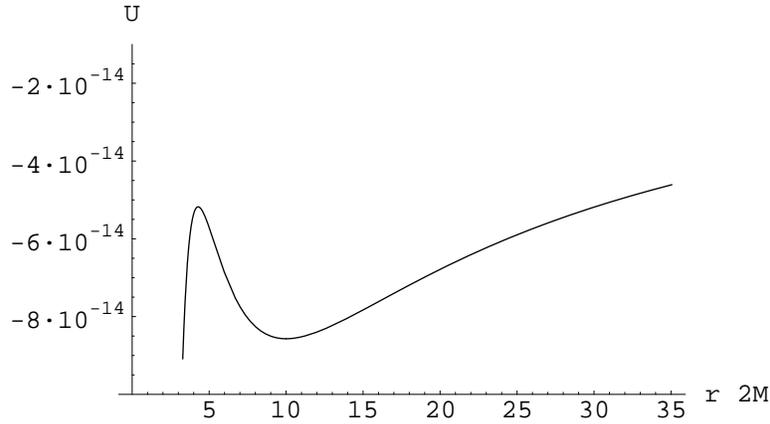}
\caption{$U_{eff}(r),\ \beta=0.4.$}
\end{figure}

From Eq.~(\ref{ueff}), we see that at the asymptotic value
$r=\infty$, the potential $U_{eff}(\infty)=0$ . To obtain a
homoclinic orbit, we must therefore require the condition
$U_{eff}(u_{un})<0$,
\begin{eqnarray}
&U_{eff}(u_{un})=-\frac{1}{27}(1+\beta)^2(1-2\beta)<0 \nonumber \\
\Rightarrow &0<\beta<\frac{1}{2}.
\end{eqnarray}
There could be a homoclinic orbit in this range of $\beta$, and we
will use the
 $\beta=0.4$ to illustrate the chaotic behavior be in the homclinc orbit.
Using the equations (3) and (5) we have the express for the
momentum
\begin{equation}
p_{r}=\frac{m}{f}\frac{dr}{d\tau}=\pm\frac{1}{f}\left[E^2-f\left(m^2+\frac{L^2}
{r^2}\right)\right]^{1/2}.
\end{equation}
In Figure 2, we obtain a homoclinic orbit in the $(r,p_r)$ reduced
phase space, the parameters are $m=10^{-6}, M=1$ and
$L=3.78\times10^{-6}$  for $\beta=0.4$.

Next, we add spin to the particle on this homoclinic orbit to show
that chaos may occur in the motion of a spinning particle around a
black hole.
\begin{figure}
\includegraphics{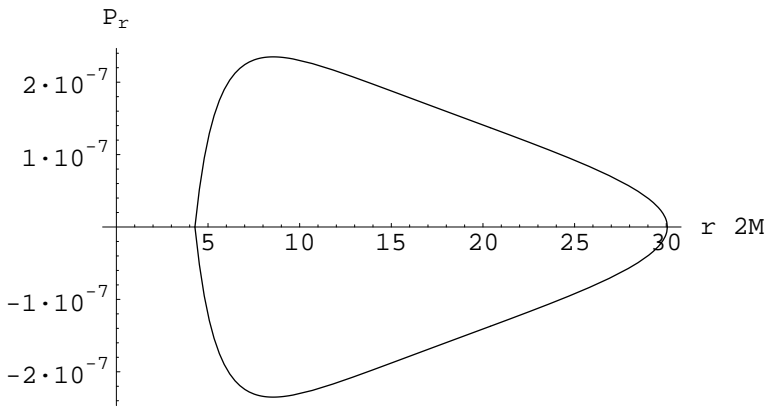}
\caption{$p_r(r)$, with $m=10^{-6},M=1,L=3.78\times10^{-6}.$}
\end{figure}
The equations of motion of a spinning test particle in a general
spacetime were first derived by Papapetrou \cite{papapetrou}.
Those are a set of equations:
\begin{eqnarray}
\frac{dx^{\mu}}{d\tau}&=&v^{\mu}, \nonumber \\
\frac{Dp^{\mu}}{D\tau}&=&-\frac{1}{2}R^{\mu}_{\
\nu\rho\sigma}v^{\nu}S^{\rho\sigma}, \nonumber \\
\frac{DS^{\mu\nu}}{D\tau}&=&p^{\mu}v^{\nu}-p^{\nu}v^{\mu},
\end{eqnarray}
$D/D\tau$ denotes a covariant derivative with respect to $\tau$.
Here $v^{\mu}$, $p^{\mu}$, and $S^{\mu\nu}$ are the 4-velocity of
the particle, the momentum, and the spin tensor, respectively. The
multipole moments of the particle higher than the mass monopole
and the spin dipole are ignored. It is called the pole-dipole
approximation.

There is a supplementary condition which gives a relation between
 $v^{\mu}$ and $p^{\mu}$  because $p^{\mu}$  is no longer parallel
to $v^{\mu}$  in this case. We adopt the following condition
\cite{dixon},
\begin{equation}
p_{\mu}S^{\mu\nu}=0.
\end{equation}
This condition is related to how to choose the center of mass in
an extended body, and we can write down the relation between
 $v^{\mu}$ and $p^{\mu}$ explicitly,
\begin{eqnarray}
v^{\mu}&=&\frac{N}{m}\left[g^{\mu\nu}p_{\nu}+\frac{1}{2m^2\Delta}
S^{\mu\nu}p_{\lambda}g^{\lambda\xi}R_{\nu\xi\rho\sigma}
S^{\rho\sigma}\right], \nonumber \\
N&=&\left[1-\frac{1}{4\Delta^2m^4}S_{\mu\nu}p_{\lambda}S_{\rho\sigma}
R^{\nu\lambda\rho\sigma}S^{\mu\alpha}p^{\beta}S^{\gamma\delta}R_{\alpha\beta\gamma\delta}
\right]^{1/2}, \nonumber \\
\Delta&=&1+\frac{1}{4m^2}R_{\alpha\beta\gamma\delta}S^{\alpha\beta}
S^{\gamma\delta}.
\end{eqnarray}
The equation shows that $p^{\mu}$ is not parallel to $v^{\nu}$ in
general due to the presence of the spin tensor. From Eq.~(11) we
get the equation of motion of the momentum $p_{\mu}$,
\begin{eqnarray}
\frac{dp_{\mu}}{d\tau}=\Gamma^{\alpha}_{\mu\nu}v^{\nu}p_{\alpha}
-\frac{1}{2}R_{\mu\nu\alpha\beta}v^{\nu}S^{\alpha\beta},
\end{eqnarray}
which will be used in the analysis of the behavior of chaotic
motions.

Because the spacetime is static and spherically symmetric, there
are two Killing vector fields, $\zeta^{\mu}_{(t)}$ and
$\zeta^{\mu}_{(\phi)}$. The constants of motion are as given in
\cite{suzuki}
\begin{eqnarray}
C&\equiv&\zeta^{\mu}p_{\mu}-\frac{1}{2}\zeta_{\mu;\nu}S^{\mu\nu},
\nonumber \\ E&=&-C_{(t)}=-p_t-\frac{M}{r^2}S^{tr}, \\
J_z&=&C_{\phi}=p_{\phi}+r^2\sin\theta\cos\theta
S^{\theta\phi}-r\sin^2\theta S^{\phi r}. \nonumber
\end{eqnarray}
$E$ and $J_z$ are interpreted as the energy of particle and the
$z$ component of the total angular momentum, respectively. The $x$
and $y$ components of the total angular momentum are also
conserved,
\begin{eqnarray}
J_x&=&-p_{\theta}\sin\phi-p_{\phi}\cot\theta\cos\phi \nonumber \\
&\quad &\ \ -r\sin\phi S^{r\theta}+r^2\sin^2\theta\cos\phi
S^{\theta\phi} +r\sin\theta\cos\theta\cos\phi S^{\phi r},
\nonumber \\ J_y&=&p_{\theta}\cos\phi-p_{\phi}\cot\theta\sin\phi
\nonumber \\ &\quad &\ \ +r\cos\phi
S^{r\theta}+r^2\sin^2\theta\sin\phi S^{\theta\phi}
+r\sin\theta\cos\theta\sin\phi S^{\phi r}.
\end{eqnarray}
Because the background is spherically symmetric, we can choose the
z axis to be in the direction of total angular momentum, then we
have the constraint equations of the spin tensor,
\begin{eqnarray}
S^{\theta\phi}&=&\frac{J}{r^2}\cot\theta, \nonumber \\
S^{r\theta}&=&-\frac{p_{\theta}}{r}, \nonumber \\ S^{\phi
r}&=&\frac{1}{r}\left(-J+\frac{p_{\phi}}{\sin^2\theta}\right).
\end{eqnarray}
From the constraint equation (12), we have the time-space
components of the spin tensor,
\begin{eqnarray}
S^{tr}&=&-\frac{1}{rp_t}\left(p^2_{\theta}+\frac{p^2_{\phi}}{\sin^2\theta}
-Jp_{\phi}\right), \nonumber \\
S^{t\theta}&=&\frac{1}{rp_t}\left(p_rp_{\theta}+\frac{J}{r}
p_{\phi}\cot\theta\right), \nonumber \\
S^{t\phi}&=&-\frac{1}{rp_t}\left(Jp_r-\frac{p_rp_{\phi}}{\sin^2\theta}
+\frac{J}{r}p_{\theta}\cot\theta\right).
\end{eqnarray}

Here we treat the spin tensor as a perturbation. Suppose that
without these perturbative terms, the particle is originally in a
homoclinic orbit on the equatorial plane $\theta=\pi/2$. When spin
terms are turned on, we write
\begin{eqnarray}
\theta&=&\frac{\pi}{2}-\delta\theta.
\end{eqnarray}
Using $\delta\theta$ to qualify the degree of perturbation, we
suppose that
\begin{eqnarray}
p_{\theta}&=&\delta p_{\theta}+O(\delta\theta^2)\nonumber\\
p_{\phi}&=&J+\delta p_{\phi}+O(\delta\theta^3)
\end{eqnarray}
where $\delta p_{\theta}\sim O(\delta\theta)$ and $\delta
p_{\phi}\sim O(\delta\theta^2)$. With these we can expand the
components of the spin tensor in Eqs.~(17) and (18).
\begin{eqnarray}
S^{\theta\phi}&=&\frac{J}{r^2}\delta\theta+O(\delta\theta^2),
\nonumber \\ S^{r\theta}&=&-\frac{\delta
p_{\theta}}{r}+O(\delta\theta^2), \nonumber
\\ S^{\phi r}&=&\frac{1}{r}(J\delta\theta^2+\delta
p_{\phi})+O(\delta\theta^3), \nonumber \\
S^{tr}&=&\frac{1}{rE}\left(\delta
p^2_{\theta}+J^2\delta\theta^2+J\delta p_{\phi}
\right)+O(\delta\theta^3), \nonumber \\
S^{t\theta}&=&-\frac{1}{rE}\left(p_r\delta
p_{\theta}+\frac{J^2}{r}\delta\theta\right) +O(\delta\theta^2),
\nonumber \\
S^{t\phi}&=&\frac{1}{rE}\left(-Jp_r\delta\theta^2+\frac{J}{r}\delta
p_{\theta}\delta\theta -p_r\delta
p_{\phi}\right)+O(\delta\theta^3).
\end{eqnarray}
We have shown explicitly only the lowest order term in each spin
tensor component in these equations.

Similarly, we also expand the 4-velocity $v^{\mu}$ and the
momentum of the particle $p_{\mu}$ in Eqs.~(13) and (14). In
particular, we have
\begin{eqnarray}
\frac{d\phi}{d\tau}&=&\frac{J}{mr^{2}}
+\frac{J}{mr^{2}}\left(1-\frac{3MJ^{2}}{m^{2}r^{3}}\right)
\delta\theta^{2}+\frac{1}{mr^{2}}\delta p_{\phi}
+O(\delta\theta^{3}).
\end{eqnarray}
Using this equation we change the parameter of the equations of
motion from $\tau$ to $\phi$. The main equations of motion of the
reduced phase space of the radial motion can then be written as
\begin{eqnarray}
\frac{dr}{d\phi}&=&\frac{1}{J}\left(1-\frac{2M}{r}\right)r^{2}p_r
-\frac{1}{J}\left(1-\frac{3MJ^2}{m^{2}r^{3}}\right)\left(1-\frac{2M}{r}\right)
r^{2}p_{r}\delta\theta^2
 \nonumber
\\ &\quad&-\frac{3MJ}{m^{2}r^{2}}\delta p_{\theta}\delta\theta
-\frac{1}{J^{2}}\left(1-\frac{2M}{r}\right)r^{2}p_r\delta p_{\phi}
+O(\delta\theta^3),\\
\frac{dp_r}{d\phi}&=&-\frac{ME^2}{J}\left(1-\frac{2M}{r}\right)^{-2}
-\frac{M}{J}p^2_r+\frac{J}{r}\nonumber \\
&\quad&+\left[\frac{M}{J}\left(1-\frac{3MJ^{2}}{m^{2}r^{3}}\right)
p_{r}^{2}-\frac{3MJ}{r^{2}}\left(1-\frac{MJ^{2}}{m^{2}r^{3}}\right)
\left(1-\frac{2M}{r}\right)^{-1}\right.\nonumber\\ &\quad&\ \
+\left.\frac{ME^{2}}{J}\left(1-\frac{3MJ^{2}}{m^{2}r^{3}}\right)
\left(1-\frac{2M}{r}\right)^{-2}
-\frac{2M^{2}J}{r^{3}}\left(1-\frac{2M}{r}\right)^{-2}\right]
\delta\theta^{2}\nonumber\\
&\quad&+\left[\frac{6M^{2}J}{m^{2}r^{4}}\left(1-\frac{2M}{J}\right)^{-1}
p_{r}\right]\delta p_{\theta}\delta\theta\nonumber\\
&\quad&+\left[\frac{1}{Jr}-\frac{3M}{Jr^{2}}\left(1-\frac{2M}{r}\right)^{-1}
-\frac{2M^{2}}{Jr^{3}}\left(1-\frac{2M}{r}\right)^{-2}\right]\delta
p_{\theta}^{2}\nonumber\\
&\quad&+\left[\frac{1}{r}+\frac{M}{J^{2}}p_{r}^{2}-\frac{3M}{r^{2}}
\left(1-\frac{2M}{r}\right)^{-1}\right.\nonumber\\ &\quad&\ \
+\left.\frac{ME^{2}}{J^{2}}\left(1-\frac{2M}{r}\right)^{-2}
-\frac{2M^{2}}{r^{3}}\left(1-\frac{2M}{r}\right)^{-2}\right]\delta
p_{\phi}\nonumber\\ &\quad&+O(\delta\theta^3).
\end{eqnarray}

\begin{figure}
\includegraphics{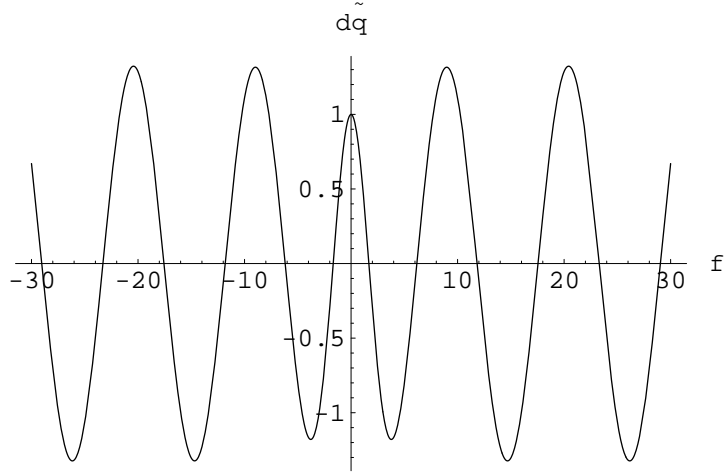}
\caption{$\delta\tilde{\theta}(\phi)$}
\end{figure}
\begin{figure}
\includegraphics{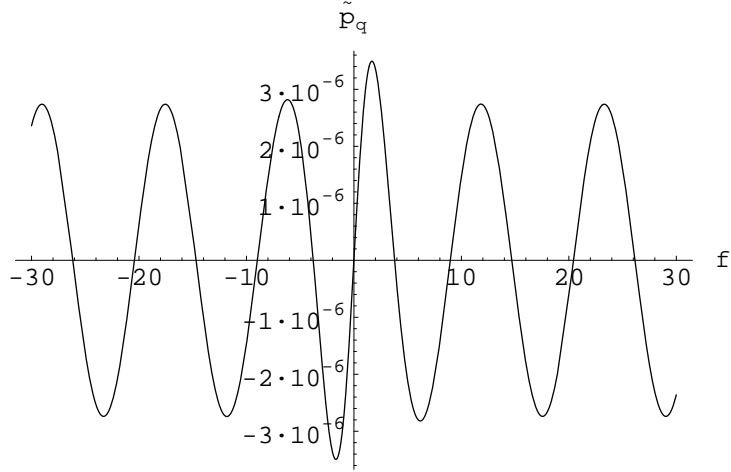}
\caption{$\tilde{p}_{\theta}(\phi).$}
\end{figure}
\begin{figure}
\includegraphics{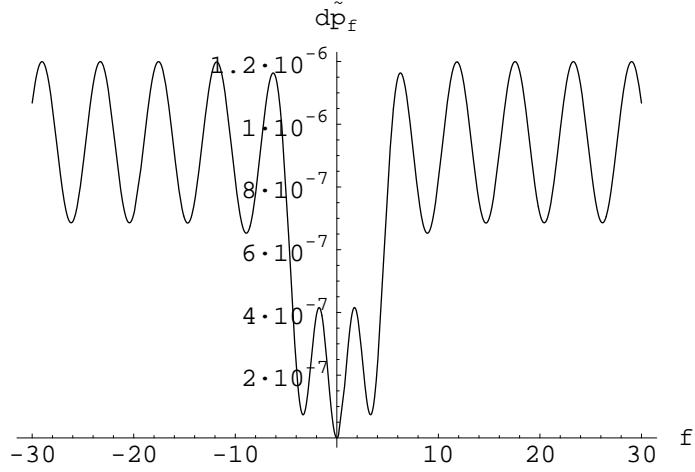}
\caption{$\delta\tilde{p}_{\phi}(\phi).$}
\end{figure}
\begin{figure}
\includegraphics{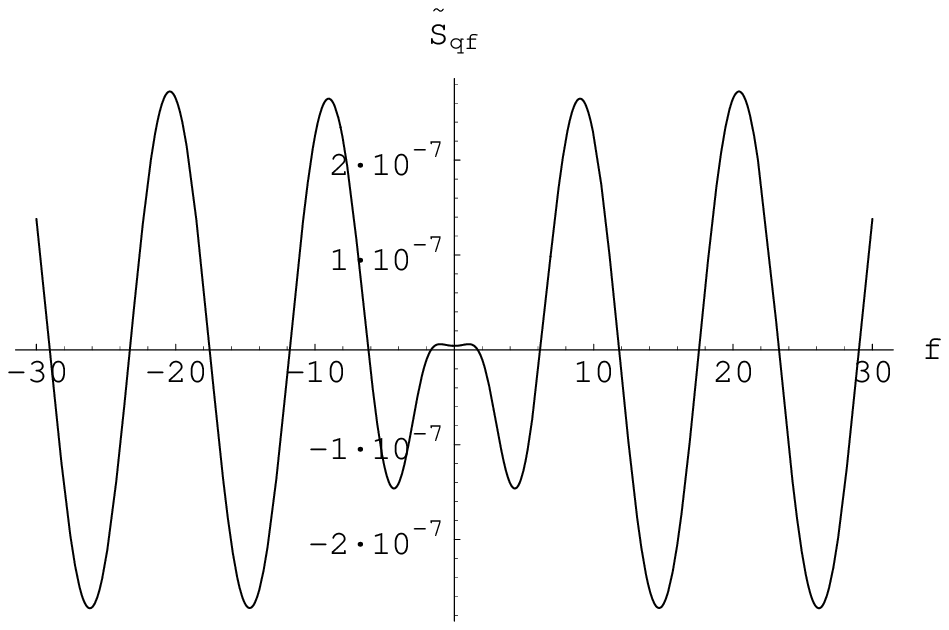}
\caption{$\tilde{S}^{\theta\phi}(\phi).$}
\end{figure}
\begin{figure}
\includegraphics{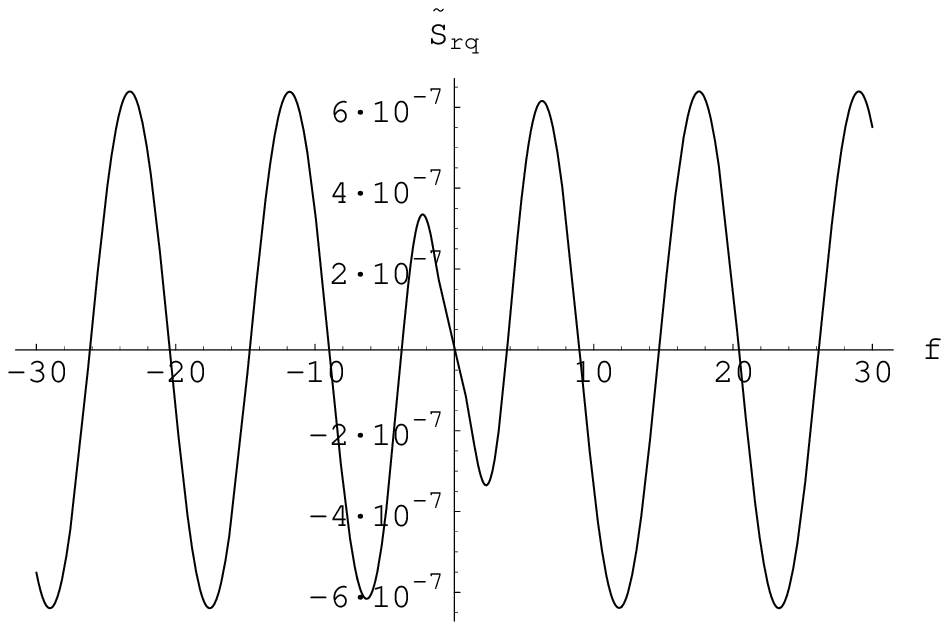}
\caption{$\tilde{S}^{r\theta}(\phi).$}
\end{figure}
\begin{figure}
\includegraphics{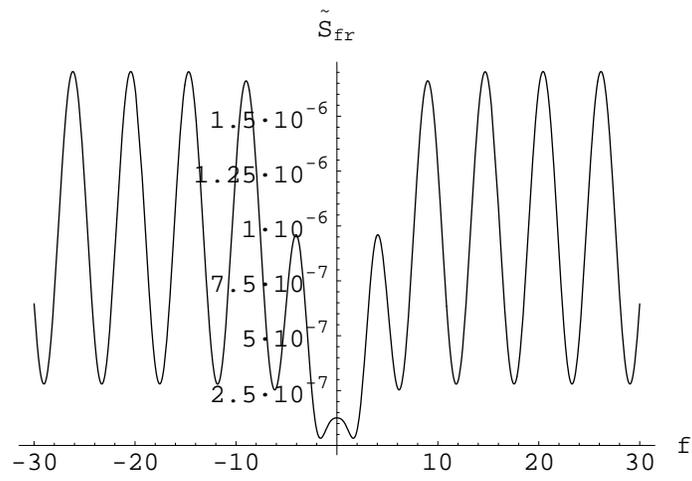}
\caption{$\tilde{S}^{\phi r}(\phi).$}
\end{figure}

There are still the other three equations of motion from Eqs.~(13)
and (14),
\begin{eqnarray}
\frac{d\delta\theta}{d\phi}&=&-\frac{1}{J}\delta
p_{\theta}+O(\delta\theta^3),
 \nonumber \\
\frac{d\delta p_{\theta}}{d\phi}&=&J\left(1-\frac{3M}{r}\right)
\delta\theta+O(\delta\theta^2),\\ \frac{d\delta
p_{\phi}}{d\phi}&=&\left[-2M+\frac{3MJ^2}{m^{2}r^{2}}
\left(1-\frac{2M}{r}\right)\right]p_r\delta\theta^2
+\frac{3M}{r}\left(1-\frac{J^2}{m^{2}r^{2}}\right)\delta
p_{\theta}\delta\theta\nonumber\\ &\quad&-\frac{2M}{J}p_r\delta
p_{\phi}+O(\delta\theta^3). \nonumber
\end{eqnarray}
These equations are used to calculate the three perturbations
$\delta\theta$, $\delta p_{\theta}$, and $\delta p_{\phi}$. We
need these quantities as inputs in the main equations of motion,
Eqs.~(23) and (24). Since we want only the lowest order terms of
these perturbations, we can take $r$ and $p_{r}$ in these
equations to be in their zeroth order, that is, corresponding to
the unperturbed homoclinic orbit. Moreover, if we define
$\epsilon\sim O(\delta\theta)$, or
$\delta\theta=\epsilon\delta\tilde{\theta}$, we can write $\delta
p_{\theta}=\epsilon\delta\tilde{p}_{\theta}$ and $\delta
p_{\phi}=\epsilon^{2}\delta\tilde{p}_{\phi}$, with
$\delta\tilde{\theta}$, $\delta\tilde{p}_{\theta}$, and
$\delta\tilde{p}_{\phi}$ all of $O(1)$. Then the solutions with
the initial conditions,
\begin{equation}
\delta\tilde{\theta}(0)=1,\quad
\delta\tilde{p}_{\theta}(0)=\delta\tilde{p}_{\phi}(0)=0,
\end{equation}
are plotted in Figs.~3-5, and the corresponding spin tensor
components are shown in Figs.~6-8. We find them all to be
oscillatory. Note that the choice of the initial conditions is
quite arbitrary. We have checked the behaviors of those quantities
for other sets of initial conditions like
$\delta\tilde{\theta}(0)=0$; $\delta\tilde{p}_{\theta}(0)=1$; $
\delta\tilde{p}_{\phi}(0)=0$, and $\delta\tilde{\theta}(0)=0$; $
\delta\tilde{p}_{\theta}(0)=0$; $\delta\tilde{p}_{\phi}(0)=1$ .
The results are also oscillatory as time evolves.

We have illustrated that external oscillatory perturbative spin
terms may induce chaotic behaviors in simple systems \cite{cho} by
calculating the Melnikov function \cite{melnikov,holmes}. Now in
the case of a spinning particle in the Schwarzschild black hole
spacetime, the Melnikov function,
\begin{eqnarray}
M(\phi_0)&=&\int^{\infty}_{-\infty}d\phi\left\{-\left[-\frac{ME^2}{J}
\left(1-\frac{2M}{r}\right)^{-2}-\frac{M}{J}p^2_{r}+\frac{J}{r}\right]g_1
+\left[\frac{1}{J}\left(1-\frac{2M}{r}\right)r^{2}p_r\right]g_2\right\}.\nonumber\\ \\
g_1&=&-\frac{1}{J}\left(1-\frac{3MJ^2}{m^{2}r^{3}}\right)\left(1-\frac{2M}{r}\right)
r^{2}p_{r}\delta\theta^2-\frac{3MJ}{m^{2}r^{2}}\delta
p_{\theta}\delta\theta
 \nonumber
\\ &\quad&
-\frac{1}{J^{2}}\left(1-\frac{2M}{r}\right)r^{2}p_r\delta
p_{\phi}\nonumber\\
g_2&=&\left[\frac{M}{J}\left(1-\frac{3MJ^{2}}{m^{2}r^{3}}\right)
p_{r}^{2}-\frac{3MJ}{r^{2}}\left(1-\frac{MJ^{2}}{m^{2}r^{3}}\right)
\left(1-\frac{2M}{r}\right)^{-1}\right.\nonumber\\ &\quad&\ \
+\left.\frac{ME^{2}}{J}\left(1-\frac{3MJ^{2}}{m^{2}r^{3}}\right)
\left(1-\frac{2M}{r}\right)^{-2}
-\frac{2M^{2}J}{r^{3}}\left(1-\frac{2M}{r}\right)^{-2}\right]
\delta\theta^{2}\nonumber\\
&\quad&+\left[\frac{6M^{2}J}{m^{2}r^{4}}\left(1-\frac{2M}{J}\right)^{-1}
p_{r}\right]\delta p_{\theta}\delta\theta\nonumber\\
&\quad&+\left[\frac{1}{Jr}-\frac{3M}{Jr^{2}}\left(1-\frac{2M}{r}\right)^{-1}
-\frac{2M^{2}}{Jr^{3}}\left(1-\frac{2M}{r}\right)^{-2}\right]\delta
p_{\theta}^{2}\nonumber\\
&\quad&+\left[\frac{1}{r}+\frac{M}{J^{2}}p_{r}^{2}-\frac{3M}{r^{2}}
\left(1-\frac{2M}{r}\right)^{-1}\right.\nonumber\\ &\quad&\ \
+\left.\frac{ME^{2}}{J^{2}}\left(1-\frac{2M}{r}\right)^{-2}
-\frac{2M^{2}}{r^{3}}\left(1-\frac{2M}{r}\right)^{-2}\right]\delta
p_{\phi}\nonumber
\end{eqnarray}
Here $r$ and $p_{r}$ are again in their zeroth order corresponding
to the unperturbed homoclinic orbit, with
$\phi\rightarrow\phi+\phi_{0}$, where $\phi_0$ is used to
parametrize the location along the unperturbed homoclinic orbit.
Moreover, $g_1$ and $g_2$ involve only the oscillatory
perturbative terms. The Melnikov function measures the transveral
distance on the Poincar$\acute{e}$ section between the perturbed stable and
unstable orbits emanating from the unstable fixed point. The
infinite number of discrete zeros of the Melnikov function as
shown in Fig.~9 indicates that the two orbits entangle with each
other, and this indicates the occurrence of chaotic behavior for
the perturbed system.

In addition, we have also calculated the Melnikov functions for
the parameter $\beta=0.1$, $0.2$, and $0.3$. Similar results as
shown in Fig.~9 with $\beta=0.4$ are obtained. These values of
$\beta$ satisfy the condition, $0<\beta<0.5$, for which
unperturbed homoclinic orbits exist. We have thus proved the onset
of chaotic motion for spinning particles near the homoclinic orbit
in the Schwarzschild black hole spacetime , even with small spin
terms. This is complementary to the results obtained by Suzuki and
Maeda \cite{suzuki}, in which chaotic behavior is detected
numerically by calculating the Lyapunov exponent in the case of
large spin terms. It is interesting to see if our method can be
extended to consider other parts of the phase space and cases with
finite spin terms. We plan to explore that in the future.

\begin{figure}
\includegraphics{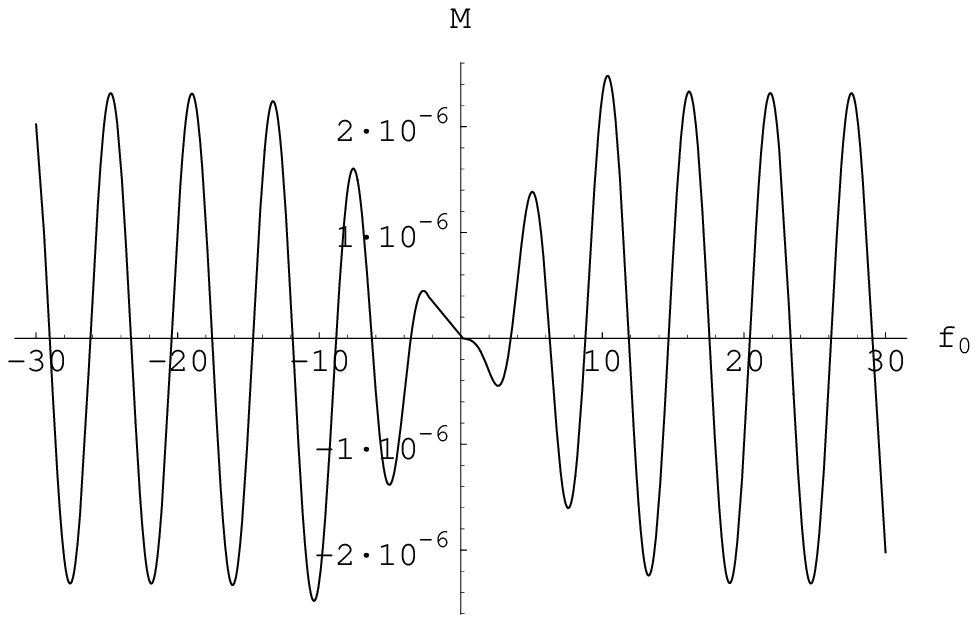}
\caption{$M(\phi_0).$}
\end{figure}

This work was supported by the National Science Council of the
Republic of China under contract number NSC 92-2112-M-032-010.

\end{document}